\documentclass[draft]{agujournal2019}
\usepackage{url} %this package should fix any errors with URLs in %refs.
\usepackage{lineno}
\usepackage[inline]{trackchanges} %for better track changes. finalnew option will compile documfent with changes incorporated.
\usepackage{soul}
\usepackage{amsmath}
\usepackage{amssymb}
\usepackage{lmodern}
\usepackage{float}
\draftfalse

\journalname{Journal of Advances in Modeling Earth Systems (JAMES)}

\begin{document}
\title{Understanding Perturbed Parameter Ensemble Sensitivities Using A Contrastive Learning Approach}
%\title{Using Contrastive Learning to Understand Parametric Uncertainties/ to calibrate Toward Calibrated Physics Ensembles}

%and Generate Calibrated Physics Ensembles
%toward calibrated physics ensemble

%%%%%%%%%%%%%%%%%%%%%%%%%%%%%%%%%%%%%%%%%%%%%%%
%
%  AUTHORS AND AFFILIATIONS
%
%%%%%%%%%%%%%%%%%%%%%%%%%%%%%%%%%%%%%%%%%%%%%%%

% Authors are individuals who have significantly contributed to the
% research and preparation of the article. Group authors are allowed, if
% each author in the group is separately identified in an appendix.)

% List authors by first name or initial followed by last name and
% separated by commas. Use \affil{} to number affiliations, and
% \thanks{} for author notes.
% Additional author notes should be indicated with \thanks{} (for
% example, for current addresses).

\authors{Da Fan \affil{1,2},
        David John Gagne II \affil{3},
        Gregory S Elsaesser\affil{4,5},
        Brian Medeiros \affil{3},
        Addisu G Semie \affil{1,2},
        Qingyuan Yang\affil{1,2},
        Akila Sampath\affil{1},
        Subashree Venkatasubramanian\affil{1},
     }

\affiliation{1}{Learning the Earth with Artificial Intelligence and Physics (LEAP) National Science Foundation (NSF) Science and Technology Center, Columbia University, New York, NY, USA}
\affiliation{2}{Department of Earth and Environmental Engineering, Columbia University, New York, NY, USA}
\affiliation{3}{NSF National Center for Atmospheric Research, Boulder, CO, USA}
\affiliation{4}{NASA Goddard Institute for Space Studies, NY, USA}
\affiliation{5}{Department of Applied Physics and Applied Mathematics, Columbia University, New York, NY, USA}
%\affiliation{=number=}{=Affiliation Address=}
%(repeat as many times as is necessary)

% Corresponding author mailing address and e-mail address:

% (include name and email addresses of the corresponding author.  More
% than one corresponding author is allowed in this LaTeX file and for
% publication; but only one corresponding author is allowed in our
% editorial system.)

% Example: \correspondingauthor{First and Last Name}{email@address.edu}

\correspondingauthor{Da Fan}{df2944@columbia.edu}

%%%%%%%%%%%%%%%%%%%%%%%%%%%%%%%%%%%%%%%%%%%%%%%
% KEY POINTS
%%%%%%%%%%%%%%%%%%%%%%%%%%%%%%%%%%%%%%%%%%%%%%%
%  List up to three key points (at least one is required)
%  Key Points summarize the main points and conclusions of the article
%  Each must be 140 characters or fewer with no special characters or punctuation and must be complete sentences

% Example:
% \begin{keypoints}
% \item	List up to three key points (at least one is required)
% \item	Key Points summarize the main points and conclusions of the article
% \item	Each must be 140 characters or fewer with no special characters or punctuation and must be complete sentences
% \end{keypoints}

\begin{itemize}
\item Contrastive learning separates two perturbed parameter ensembles (PPEs) in embedding space while preserving seasonal variability and ensemble spread due to parameter perturbations.
\item Attribution of embedding distances between PPEs and observations highlights subtropical low-cloud regions, storm tracks regions, and tropical convection regions.
\item Correlations between regional attributions and perturbed parameters indicate that embedding distances encode regional sensitivities of cloud and radiative fields to physics parameters.
\end{itemize}

%%%%%%%%%%%%%%%%%%%%%%%%%%%%%%%%%%%%%%%%%%%%%%%
%
%  ABSTRACT and PLAIN LANGUAGE SUMMARY
%
% A good Abstract will begin with a short description of the problem
% being addressed, briefly describe the new data or analyses, then
% briefly states the main conclusion(s) and how they are supported and
% uncertainties.

% The Plain Language Summary should be written for a broad audience,
% including journalists and the science-interested public, that will not have 
% a background in your field.
%
% A Plain Language Summary is required in GRL, JGR: Planets, JGR: Biogeosciences,
% JGR: Oceans, G-Cubed, Reviews of Geophysics, and JAMES.
% see http://sharingscience.agu.org/creating-plain-language-summary/)
%
%%%%%%%%%%%%%%%%%%%%%%%%%%%%%%%%%%%%%%%%%%%%%%%

%% \begin{abstract} starts the second page

\begin{abstract}
Perturbed parameter ensembles (PPEs) reveal how physics parameters affect climate simulations, but interpreting parameter sensitivities across multivariate, spatially structured outputs remains challenging, particularly when calibrating models against observations. We develop an explainable contrastive learning model that maps 5 monthly cloud and radiation fields into a shared representation space. We train the model on the fields of two 100-member Community Atmosphere Model version 6 (CAM6) PPEs, spanning 34 parameters, that only differ in the warm rain microphysics scheme: KK2000, the default bulk microphysics scheme, and TAU-ML, a neural network emulator of a bin microphysics scheme. The learned representations separates two PPEs with over 94\% linear classification accuracy while preserving the seasonal variability and ensemble spread due to parameter perturbations. In the shared representation space, the representations of satellite observations occupy the same low-dimensional manifold as the PPEs but are displaced from them most strongly during boreal spring and autumn. TAU-ML PPE has a lower distance to observations compared to KK2000 in the representation space. Integrated Gradients attributions highlights the contributions in subtropical low-cloud regions, Northern and Southern Hemisphere storm track regions, and tropical convection regions to differences between PPEs and observations. Regional attributions correlate most strongly with parameters associated with cloud microphysics, boundary layer turbulence, and deep convection. These results demonstrate that explainable representations of climate fields can attribute model differences to specific variables, regions, seasons, and physical parameters.
\end{abstract}

\section*{Plain Language Summary}
Climate models cannot directly resolve many cloud and rainfall processes, so these processes are represented using simplified equations containing uncertain choices. Testing many choices produces a large collection of simulations, but comparing all of their spatial patterns and variables with observations is difficult. We trained a machine-learning method to organize monthly cloud-water and radiation fields from two groups of 100 climate simulations. The groups used the same uncertain parameter values but different representations of how cloud droplets combine to form rain. The learned organization preserved the seasonal cycle and the spread caused by parameter choices while distinguishing the two rain-formation methods. Both simulation groups differed more from observations than from each other, although the more detailed rain representation was generally closer to observations. Cloud liquid water and reflected sunlight explained most of the differences, especially over marine low-cloud regions and midlatitude storm tracks. The method also connected these regional differences to parameters controlling cloud formation, atmospheric turbulence, and deep convection. This approach can help model developers determine not only which simulations agree better with observations, but also which physical processes and regions produce that agreement.

%%%%%%%%%%%%%%%%%%%%%%%%%%%%%%%%%%%%%%%%%%%%%%%
%
%  BODY TEXT
%
%%%%%%%%%%%%%%%%%%%%%%%%%%%%%%%%%%%%%%%%%%%%%%%

%%% Suggested section heads:

\section{Introduction}

Earth system models (ESMs) connect a growing number of Earth system component models to enable higher fidelity simulations of past climate variability and future climate change. At climate timescales, uncertainties and biases arise more from the complex web of modeling choices made when configuring each ESM run. Many microphysics, turbulence, convection, and aerosol processes occur at scales smaller than the grid spacing of current ESMs and must therefore be represented through sub-grid parameterizations \cite{Flato2011, SchmidtSherwood2015, Gettelman2022}. These parameterizations contain uncertain parameters and structural assumptions that influence the simulated cloud, precipitation, radiation, and ultimately climate sensitivities \cite{Sherwood2020, Tapiador2019}. As models continue to increase in complexity and resolution, we need more scalable and systematic ways to evaluate and constrain uncertainty in physical parameters and structural bias \cite{Morrison2020, Fletcher2022}.

Climate model development has traditionally relied on hand tuning to reduce biases in energy and hydrological cycle metrics. Although this approach has produced substantial improvements, it is time-consuming, subjective, and increasingly difficult as the number of uncertain parameters, model outputs, and observational constraints grows \cite{Hourdin2017, Mauritsen2012, Schmidt2017, Mignot2021}. Moreover, multiple parameter combinations can produce similarly acceptable mean climate states, a condition often described as equifinality \cite{BevenFreer2001, Her2019}. Agreement with a small set of scalar metrics also does not guarantee that the model correctly represents the underlying physical processes. A model may compensate for errors across variables, regions, or seasons, producing apparently improved global metrics while retaining important structural deficiencies \cite{Hourdin2017, Elsaesser2025}.

Perturbed parameter ensembles (PPEs) provide a powerful framework for exploring these uncertainties and deficiencies. By systematically perturbing parameters in cloud microphysics, boundary layer turbulence, shallow and deep convection, and aerosol schemes, PPEs reveal how uncertain parameter choices affect simulated climate \cite{Sexton2021, Eidhammer2024, Duffy2024}. They also provide a basis for generating calibrated physics ensembles (CPEs), in which parameter combinations are selected to improve agreement with observations while retaining physically plausible solutions \cite{Elsaesser2025}. However, interpreting PPEs remains difficult because model output is high-dimensional, multivariate, spatially structured, and seasonally dependent. Differences among simulations, or between simulations and observations, may reflect parameter uncertainty, structural error, internal variability, observational uncertainty, or compensating biases among fields \cite{Braverman2021, Yang2024, Elsaesser2025}.

A key challenge is therefore to develop diagnostics that compare models and observations in a compact but physically meaningful space. Such diagnostics should preserve regional and seasonal structure, represent interactions among variables, distinguish model configurations, quantify discrepancies between the model and observations, and identify which variables, regions, and parameters control those differences. This need is especially important for CPEs, where the goal is not only to reduce a scalar error metric but also to determine whether calibration improves the cloud and radiation processes \cite{Hourdin2023, WatsonParris2021, Yarger2024, Elsaesser2025}.

Machine learning offers new tools for addressing this problem. Contrastive learning is a self-supervised representation learning approach that trains a neural network to place similar samples closer and dissimilar samples farther apart in a compact embedding space \cite{Oord2019,Hjelm2019,Chen2020,He2020}. Related developments have shown that learned representations can capture similarities and differences among samples without requiring a manually specified distance metric \cite{Khosla2021,Tian2020}. For climate applications, this approach is attractive because it can learn compact representations of multivariate spatial or spatiotemporal fields while retaining information relevant to downstream prediction, classification, bias correction, and physical interpretation \cite{Ballard2022,Wang2022,Lv2022,Gong2024,Chatterjee2023,Wang2025,Bailey2026}. If trained appropriately, the resulting embedding space might encode seasonal variability, ensemble spread, and differences among model configurations and observations. Distances in this space can then be interpreted as data-driven measures of similarity among climate states.

In this study, we develop a contrastive learning framework to quantify sensitivities to physical parameters and diagnose structural deficiencies in climate model simulations and support the development of parameter calibration work \cite{Hourdin2017, Mignot2021, Yang2024, Elsaesser2025}. We apply the framework to two CAM6 PPEs that differ in their warm rain microphysics schemes: one uses the default KK2000 scheme, and the other uses the TAU-ML bin scheme (described in further detail later). This paper addresses three questions specifically:
\begin{enumerate}
    \item Can contrastive learning construct a representation space that captures seasonal variability, parametric uncertainty, and structural differences among TAU-ML PPE, KK2000 PPE, and observations?
    \item Can distances in this representation space provide useful measures of parametric uncertainty and structural error?
    \item Which variables, regions, and physical parameters control the variability in the representation and physical space?
\end{enumerate}
%Because both ensembles perturb parameters across cloud microphysics, CLUBB boundary layer turbulence, shallow and deep convection, and aerosol schemes, this design allows us to examine both parametric uncertainty and structural differences associated with the warm-rain parameterization \cite{Eidhammer2024}.
% One ensemble uses the default Khairoutdinov and Kogan warm rain parameterization, hereafter KK2000, which represents auto-conversion and accretion using empirical power law relationships \cite{KhairoutdinovKogan2000}. The other uses TAU-ML, a neural network emulator of the Tel Aviv University (TAU) bin microphysics model that replaces the KK2000 warm-rain process with an emulated stochastic collection process \cite{Gettelman2021}. 

The remainder of the paper is organized as follows. Section~2  describes the CAM6 PPEs, warm rain parameterizations, and observational datasets. Section~3 introduces the contrastive learning framework, embedding evaluation methods, and integrated gradient attribution approach. Section~4 presents the learned embedding structure, seasonal embedding distances, parameter sensitivities, attribution maps, and regional parameter sensitivities. Section~5 summarizes the main conclusions and discuss their implications for parameter calibration.

\section{Data}
Here we describe the CAM6 model used (Section ~\ref{2.1}), the PPE simulations (Section ~\ref{2.2}), and observational datasets (Section ~\ref{2.3}).
\subsection{CAM6 model description}
\label{2.1}
The Community Atmosphere Model Version 6 (CAM6) is the atmospheric component of the Community Earth System Model version 2 \cite{danabasoglu2020community}. CAM6 features a two-moment stratiform cloud microphysics scheme \cite[MG2;]{gettelman2015advanced} with prognostic cloud liquid, cloud ice, rain, and snow hydrometeors. For these hydrometeors, MG2 predicts both mass and number concentration. MG2 is coupled to a unified moist turbulence scheme, Cloud Layers Unified by Binormals \cite{golaz2002pdf, Larson2002-lg}, implemented by \citeA{Bogenschutz2013-bz}. CLUBB treats boundary layer moist turbulence, and grid-scale cloud processes. 
CAM6 features a 4-mode aerosol model \cite{Liu2016-hl} with an extra mode for primary carbon and a representation of natural and anthropogenic aerosols. CAM6 also features an ensemble plume mass flux deep convection scheme \cite{Neale2008-dm, zhang1995sensitivity} with simple microphysics, a physically based mixed phase ice nucleation scheme \cite{Hoose2010-ax}, and the Rapid Radiative Transfer Model for General Circulation Models \cite[RRTMG;]{Iacono2000-jo} radiation scheme.

All CAM6 (version \texttt{6\_4\_046}) experiments use a standard $0.9^\circ \times 1.25^\circ$ (latitude and longitude) horizontal resolution with 32 vertical levels (minimum at 2 hPa) and are configured with present day climatological forcing, where Sea Surface Temperature (SSTs), sea ice, and boundary forcings are prescribed from climatological conditions centered on year 2000. 

\subsection{PPE simulations}
\label{2.2}
We leverage two 100-member PPEs using CAM6 with the only difference being the warm rain microphysics parameterizations. The two microphysics parameterizations are more completely described in \cite{Gettelman2021}, but here we repeat key information. The first warm rain parameterization is the default two-moment bulk microphysics scheme \cite[hereafter KK2000]{KhairoutdinovKogan2000}, which represents warm rain formation process by autoconversion and accretion. KK2000 defines autoconversion and accretion rate by fitting empirical power law to large eddy simulations (LES) with bin-resolved microphysics. The autoconversion rate components have been adjusted to better match observations \cite{gettelman2019high}. 

The other warm rain parameterization is the machine learning emulator of the Tel Aviv University (TAU) bin microphysical model (hereafter TAU-ML). The TAU model resolves the drop size distrbutions using 35 bins, which allows for mass transfer between bins and alleviates anomalous drop growth. The autoconversion and accretion rates in KK2000 are replaced by explicitly solving the quasistochastic collection process. However, the quasistochastic collection process is computationally expensive for practical global climate and weather model simulations. A neural network was used in \citeA{Gettelman2021} to emulate the stochastic collection tendency from cloud and rain water variables, which shows consistent results with the TAU model and a variant of the neural network \cite{Gagne2026} is used in our study as TAU-ML.

\begin{table}[H]
    \caption{Perturbed parameter names, descriptions, default values, rangesm and units for the 100 member PPE.}
    \noindent\includegraphics[width=\textwidth]{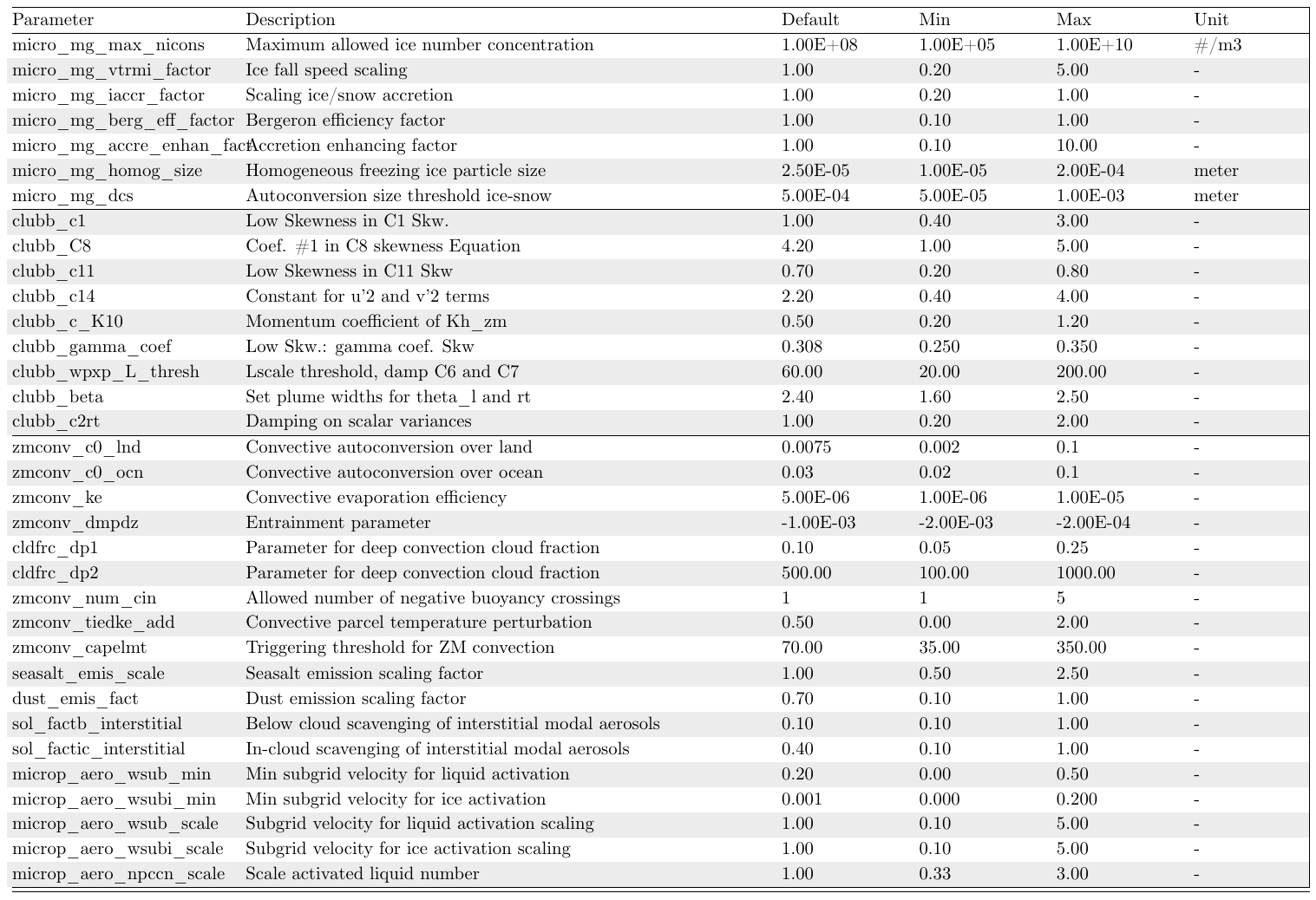}
    \label{Table1}
\end{table}

The two PPE simulations for KK2000 and TAU-ML consist of 100 ensemble members with 34 parameters (Table \ref{Table1}) sampled by Latin hypercube sampling. These parameters span the turbulence \cite<CLUBB;>{golaz2002pdf}, cloud microphysics \cite<MG2;>{gettelman2015advanced}, aerosol \cite<MAM;>{liu2012toward}, and deep convection \cite<ZM;>{zhang1995sensitivity}. Their ranges are determined with expert elicitation \cite<their Table 1,>{Eidhammer2024}. The same parameter sets are used in both PPEs, but note that the accretion parameter (\texttt{micro\allowbreak\_mg\allowbreak\_accre\allowbreak\_enhan\allowbreak\_fact}) is not activated when the TAU-ML scheme replaces KK2000. We work with the CAM6 PPE present-day simulations. Each PPE member simulation is run using near present day cyclic boundary conditions of the year 2000. The greenhouse gases and atmospheric oxidants take the average values of the 1995-2005 period, the average monthly sea surface temperatures (SSTs) during 1995-2010 are used, and the emission of aerosols and precursors is set to 1995-2005 in the simulations. Each simulation runs for a period of 3 years and the output is archived at monthly intervals for climatological fields. We focus on 5 variables, which characterize the cloud and radiative responses to changes in microphysics and parameters, of the CAM6 PPE in this work: shortwave cloud radiative effect (SWCF), long wave cloud radiative effect (LWCF), upwelling longwave flux at top of model (FLUT), net shortwave flux at top of model (FSNT), and total (cloud plus rainwater) gridbox-average liquid water path (TLWP). These variables were selected because they exhibited substantial differences compared to observations \cite[their Fig.~12]{Gettelman2021} and strong sensitivities to parameter changes in the CAM PPE \cite[their Fig.~3]{Eidhammer2024}. Four static variables, land-sea mask, surface height, latitude, and longitude, are also included in the inputs to provide geographic and surface context.

\subsection{Observations}
\label{2.3}
\begin{table}[H]
    \caption{Monthly climatological observational products used in this work.}
    \noindent\includegraphics[width=\textwidth]{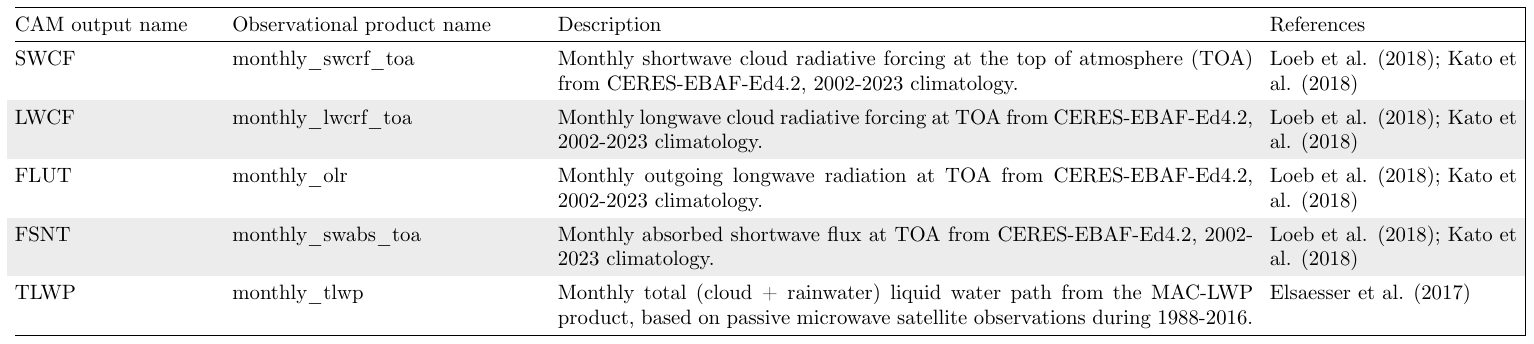}
    \label{Table2}
\end{table}

To evaluate PPE outputs, we use the observational monthly climatologies of LWCF, SWCF, FLUT, FSNT \cite{Loeb2018, Kato2018}, and TLWP (\citeA{Elsaesser2017}), derived from products in Table \ref{Table2}, as the metrics of cloud and top-of-atmosphere radiation fields. Observational uncertainty has limited impacts on the results and is therefore not considered in this study, but will be incorporated into future calibration work.

%Instead of focusing on the global and temporal averages (scalars) or the temporal averages (rasters), we propose focusing on the local regions. 
%monthly mean
%Pairing strategy
%Data split

\section{Methods}
\subsection{Data Processing}
%Table to show
The dataset comprises monthly samples from the KK2000 and TAU-ML PPEs and observational climatologies between 65.5$^\circ$S and 65.5$^\circ$N. The polar regions are excluded due to the large observational uncertainties in cloud retrievals over snow and ice covered surfaces. Each sample includes anomalies of five monthly mean variables (SWCF, LWCF, FLUT, FSNT, and TLWP) and four static fields (latitude, longitude, land-sea mask, and surface height). At each grid point, we compute the anomalies by standardizing each variable separately, using three-year mean and standard deviation for each PPE member and the mean and standard deviation across 12 monthly climatologies for observations. We exclude the first month of each PPE simulation because ensemble spread is initially reduced as members diverge from the same initial conditions. The PPE samples are split by ensemble member into training (members 1–60; 4200 samples), validation (members 61–80; 1400 samples), and testing (members 81–100; 1400 samples) subsets.

\subsection{Contrastive Learning framework}
\begin{figure*}[!th]
\centerline{\includegraphics[width=\textwidth,angle=0]{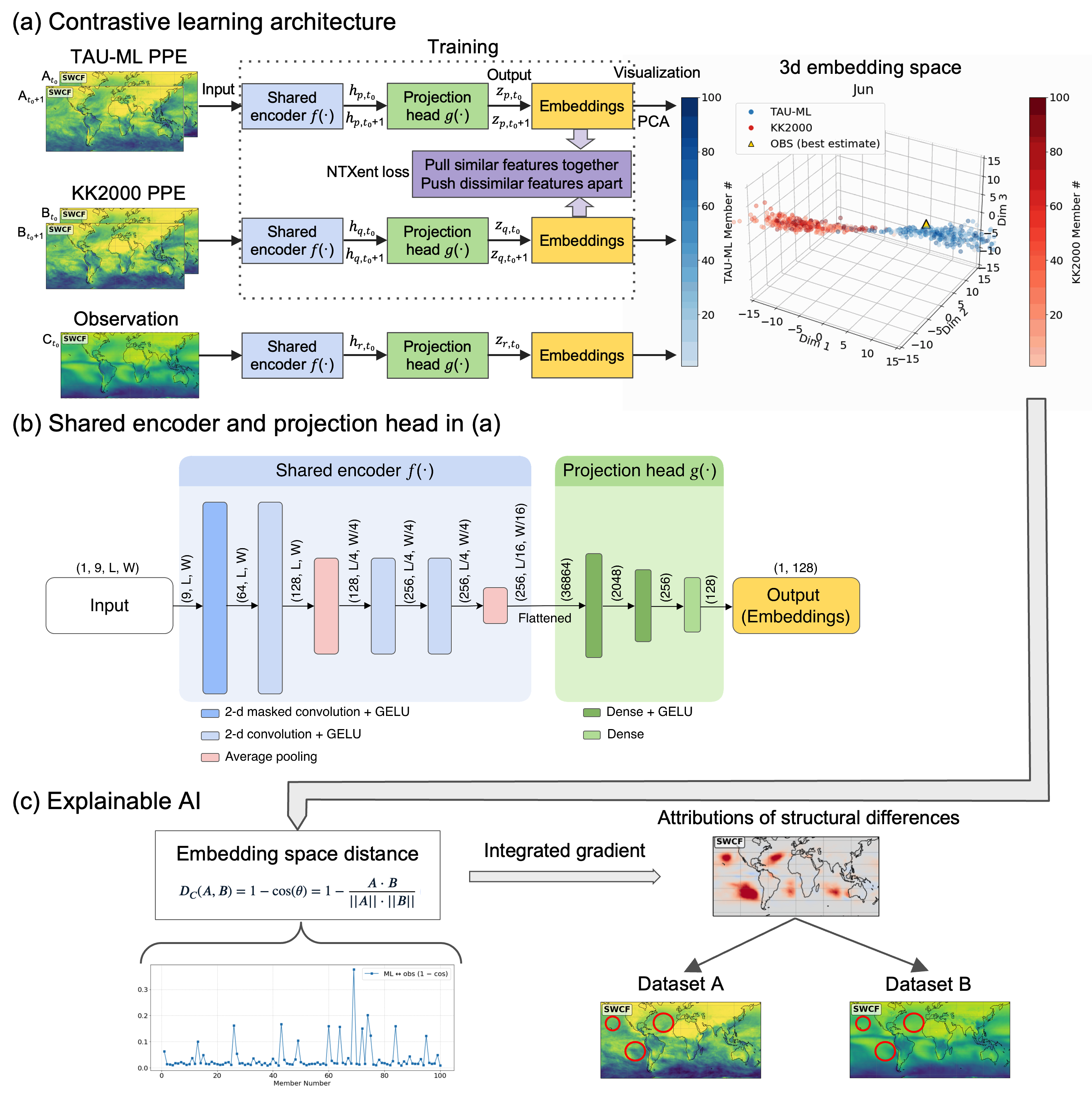}}
\caption{Schematic of the contrastive learning framework and explainable AI workflow. (a) Contrastive learning architecture showing the training and inference pipelines and a visualization of the two-dimensional projections of the learned embeddings. The model uses five atmospheric variables and four static fields; only SWCF is shown for illustration. (b) Shared neural network encoder and projection head that map input from different data sources into a common embedding space. (c) Explainable AI workflow that uses the Integrated Gradient method to attributes cosine distances between embeddings to geographic regions of the input fields.}\label{Fig1}
\end{figure*}

%(a) Contrastive learning architecture. Monthly inputs from TAU-ML and KK2000 Perturbed parameter ensembles (PPE), together with observations, are passed through a shared neural network encoder to obtain latent feature representations. During training, the NT-Xent loss pulls similar latent features of TAU-ML and KK2000 together and pushes dissimilar latent features apart in the latent space. The resulting 3-dimensional projection (via PCA) illustrates the separation between datasets in the learned feature space. (b) Structure of the neural network encoder. The model consists of convolutional and average-pooling layers followed by fully connected (dense) layers, mapping the input fields to a 128-dimensional latent feature vector. (c) Explainable AI framework. Structural differences between datasets are quantified using cosine distance in latent space. Integrated gradients are applied to attribute latent space differences to spatial regions in the input fields, highlighting areas that contribute most to discrepancies between Dataset A and Dataset B.
%The effective batch size during training is 400 samples, achieved through a multi-step training scheme (100 steps) and a mini-batch size of 4 combining both model ensembles.
%To explore climate model bias, 
%Q: should I explain what is embedding?

We develop a self-supervised contrastive learning framework (Fig. \ref{Fig1}) that maps climate model outputs and observations to low-dimensional representations in a shared embedding space. Contrastive learning \cite{Chen2020} is a discriminative approach that rewards pulling similar samples closer in the embedding space while also pushing dissimilar samples apart.

Fig. \ref{Fig1}a shows the contrastive learning model architecture. The model is trained on mini-batches, with each containing two pairs of samples: $(A_{t}, A_{t+1})$ from a TAU-ML ensemble member and $(B_{t}, B_{t+1})$ from its corresponding KK2000 member. The two members share the same perturbed physics parameters and both pairs span the same adjacent months, $t$ and $t+1$. The pairs from the same microphysics scheme, $(A_{t}, A_{t+1})$ and $(B_{t}, B_{t+1})$, are treated as positive pairs because the adjacent monthly outputs within each pair represent consecutive segments of the same dynamical trajectory and therefore expected to share large scale physical characteristics. The pairs from different schemes, $(A_{t}, B_{t})$, $(A_{t}, B_{t+1})$, $(A_{t+1}, B_{t})$, and $(A_{t+1}, B_{t+1})$, are considered as negative pairs because their samples represent segments from different dynamic trajectories generated with different schemes. In each positive pair, the two monthly samples represent distinct, partial views of the same dynamical trajectory from a single TAU-ML or KK2000 member. By contrast, the two samples in each negative pair represent partial views of different trajectories. By contrasting the embeddings of positive pairs and negative pairs, the model learns a representation that captures similarities within a dynamical trajectory and differences between trajectories.

We use a shared neural network (NN) architecture (Fig. \ref{Fig1}b), including a shared encoder $f(\cdot)$ and a projection head $g(\cdot)$, to project input data $x$ into embeddings that represent radiation and liquid water characteristics. The shared encoder comprises a channel-aware masked convolutional layer followed by three same-padding convolutional layers, all with 3×3 kernels. The masked convolution is applied to the TLWP channel to extract features only over water regions, with the output renormalized to remain comparable to the other channels. Hyperparameters were selected through manual tuning. The number of convolutional filters increases after every convolutional layer, from 64 to 128 and then to 256, with the last two convolutional layers both using 256 filters. Gaussian Error Linear Unit (GELU; \citeNP{Hendrycks2016}) activation function is applied after each convolutional layer. Two average pooling layers, each with a 3×3 kernel and a stride of 4, downsample the feature maps after every two convolutional layers. The final feature map is flattened into a one-dimensional feature vector, $h = f(x)$ and passed through the projection head. The projection head applies three dense layers with 2048, 256, and 128 units, to reduce the dimension of the embedding space. GELU activations are applied after the first two dense layers and the final dense layer generates a 128-dimensional embedding $z = g(h)$.
%($z_{p,t}, z_{p,t+1}, z_{q,t}, z_{q,t+1}$)
%Manual hyperparameter tuning
We apply the shared encoder and projection head to map inputs from TAU-ML and KK2000 datasets into a common embedding space. During training, the model is optimized using the normalized temperature-scaled cross-entropy (NT-Xent) loss, that groups embeddings of positive pairs together while distinguishing those of negative pairs. Each minibatch contains $N=2$ positive pairs and four negative pairs, for a total of $2N=4$ samples. Let
\begin{equation}
\operatorname{sim}(z_i,z_j)
=
\frac{z_i^\top z_j}
{\lVert z_i\rVert_2\lVert z_j\rVert_2}
\end{equation}
denote the cosine similarity between embeddings $z_i$ and $z_j$, equivalent to the dot product of their $\ell_2$-normalized representations. For each sample $i$, let $p(i)$ denote the other sample in its positive pair. The NT-Xent loss for a minibatch is
\begin{equation}
\mathcal{L}_{\mathrm{batch}}
=
-\frac{1}{2N}
\sum_{i=1}^{2N}
\log
\frac{
\exp\!\left[\operatorname{sim}(z_i,z_{p(i)})/\tau\right]
}{
\displaystyle
\sum_{k=1}^{2N}
\mathbb{I}_{[k\neq i]}
\exp\!\left[\operatorname{sim}(z_i,z_k)/\tau\right]
},
\end{equation}
where $\mathbb{I}_{[k\neq i]}$ equals 1 when $k\neq i$ and 0 otherwise, and $\tau$ is the temperature hyperparameter. The loss is averaged over all four samples in the minibatch, with each sample serving once as the anchor.

We trained the contrastive learning model with minibatches. The loss was computed for each minibatch and gradients were accumulated over 100 minibatches before each optimizer update, corresponding to an effective batch size of 400. We used the Adam optimizer \cite{Kingma2014} with a learning rate of 0.0001 and weight decay of $10^{-5}$ for gradient descent. Early stopping was applied when the validation loss does not improve over 10 epochs.

After training, we applied the model to extract embeddings from KK2000, TAU-ML, and observation datasets. We then applied two approaches to evaluate the robustness of embeddings. First, we trained a logistic regression model using the embeddings to separate TAU-ML and KK2000 classes. The skill of the logistic regression in separating the classses provided initial evidence that the contrastive learning procedure was learning a robust signal. Second, we applied principle component analysis (PCA) to reduce the 128-dimensional embeddings from all data sources to 3 dimensions for visualization. We then compared the projected embeddings across data sources to assess the robustness of the learned representation.

\subsection{Explainable AI}
We employed the Integrated Gradients \cite{Sundararajan2017} to explain the learned embedding space and identify differences among different datasets (Fig. \ref{Fig1}b). The Integrated Gradients method generates heatmaps that represent the relative contributions of geographic regions for climate applications \cite{Mamalakis2022}. Each heat map was normalized by its absolute maximum value and averaged across all ensemble members and months to highlight the structural differences. Integrated gradients produces similar attribution patterns to the model-agnostic SHAP method but is much faster to compute.

\section{Results}
\subsection{Embedding evaluation}
\begin{figure*}[!th]
\centerline{\includegraphics[width=\textwidth,angle=0]{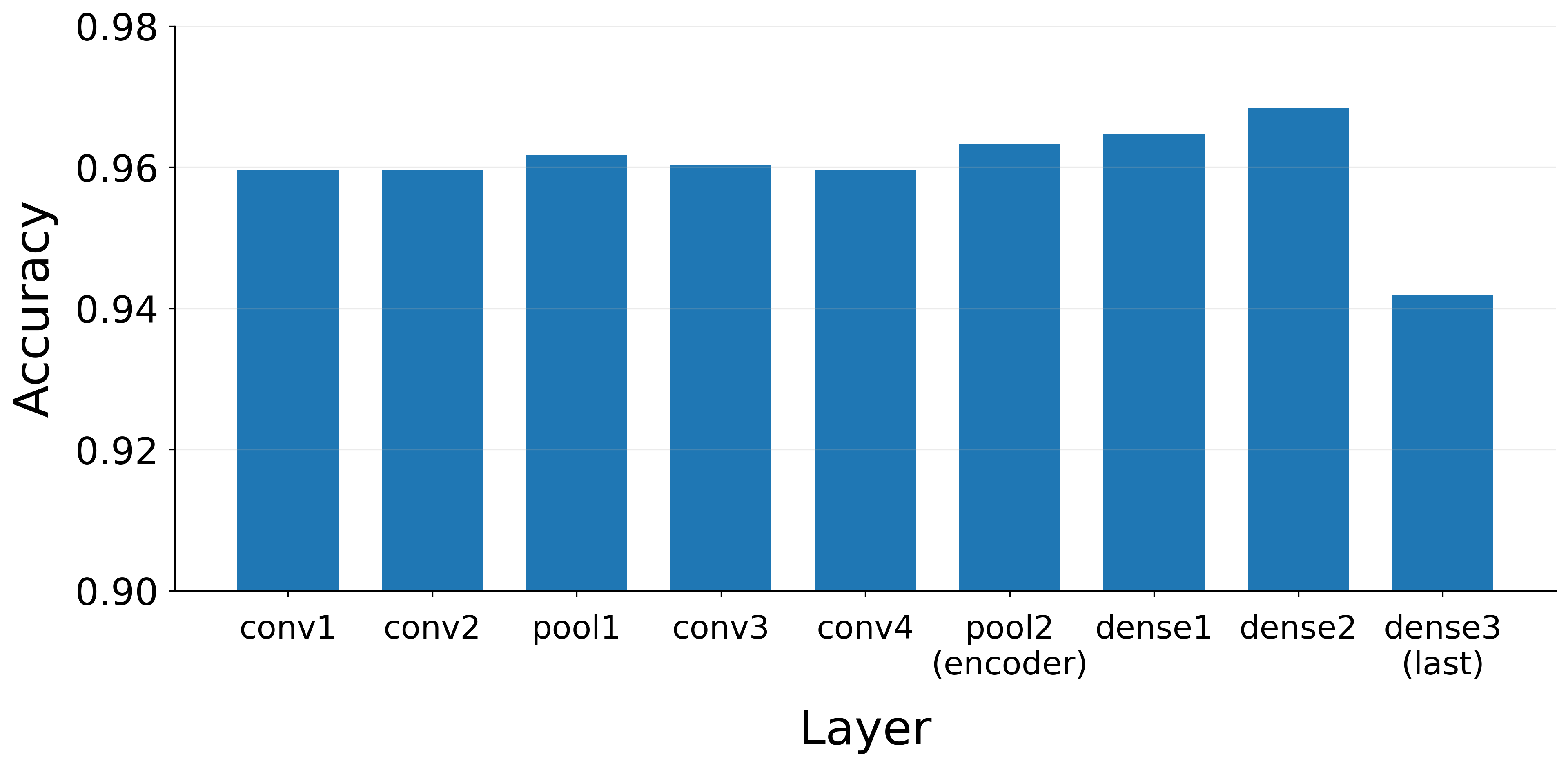}}
\caption{Linear (TAU-ML vs KK2000) classification evaluation of the learned embeddings from different layers of the contrastive learning model.}\label{Fig2}
\end{figure*}
We first evaluate whether the learned representations contain information that distinguishes KK2000 and TAU-ML samples, while also capturing variability arising from parameter perturbations and structural errors relative to observations. We trained logistic regression classifiers using embeddings extracted by each network layer from TAU-ML and KK2000 training samples and evaluated them on testing samples. Figure~\ref{Fig2} shows that embedding-based classification skill remains high across all layers, with accuracies over 0.94. The highest skill occurs at the dense2 layer, whereas the lowest skill is at the last dense3 layer. Notably, embeddings from the first conv1 layer already achieve high classification accuracy, suggesting that the earliest learned representations contain information for distinguishing between datasets. The limited skill improvement from the first conv1 layer to the deeper dense2 layer suggests that intermediate layers provide only incremental information for TAU-ML and KK2000 separation. The accuracy decrease from dense2 to dense3 may reflect the role of the final dense layer in compressing the representation into a lower-dimensional latent space optimized for the contrastive loss rather than for classification. This compression may discard some discriminative information while retaining features that better preserve embedding similarity. In the following analysis, we focus on embeddings from pool2 and dense3 as representative outputs of the encoder and projection head, respectively.

\begin{figure*}[!th]
\centerline{\includegraphics[width=\textwidth,angle=0]{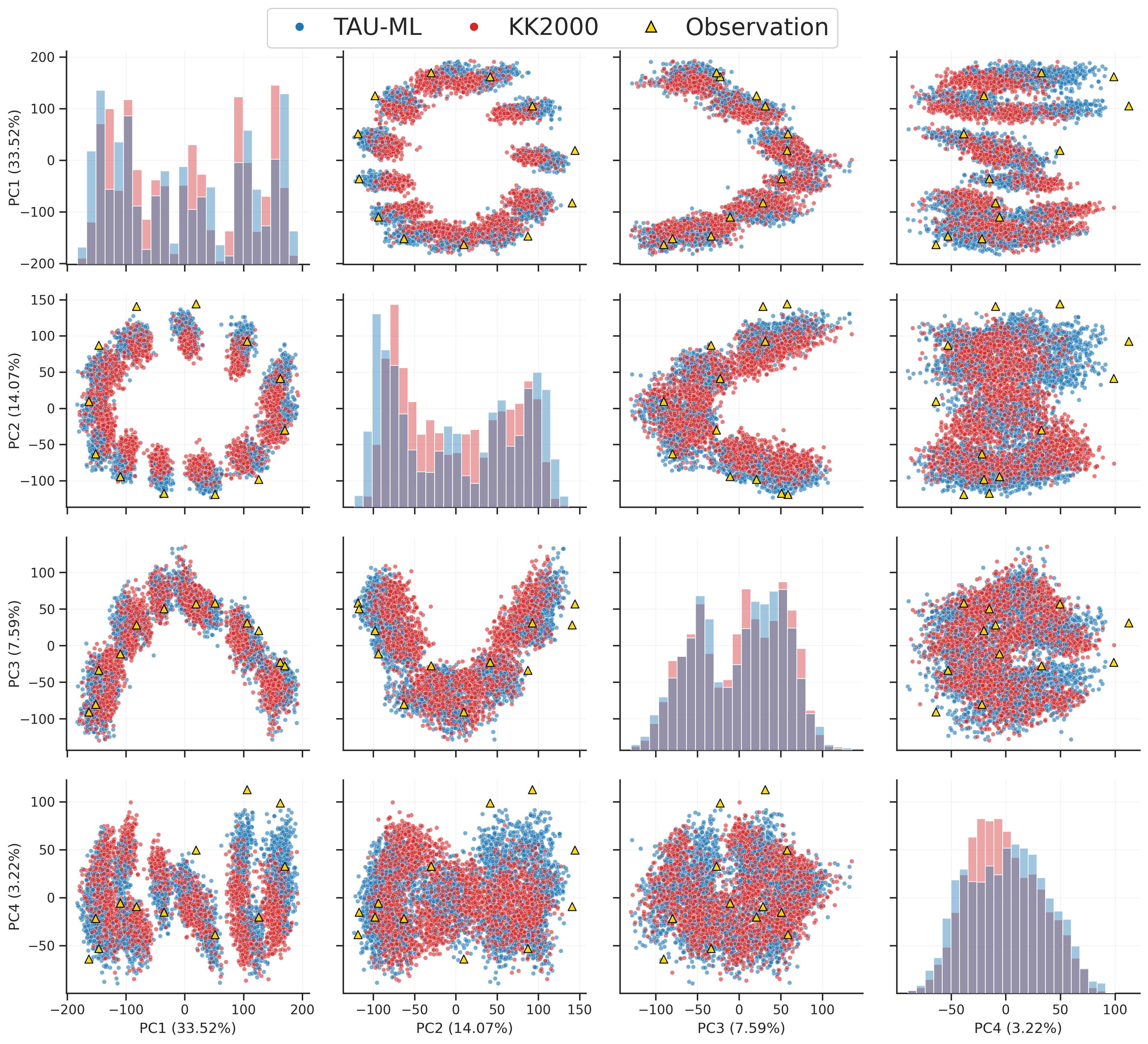}}
\caption{Pairwise two-dimensional (2D) PCA projections of the learned encoder (pool2) embeddings for TAU-ML, KK2000, and observations. Diagonal panels show distributions for each principal component (PC), while off-diagonal panels show pairwise projections among PC1, PC2, PC3, and PC4. Blue circles denote TAU-ML samples, red circles denote KK2000 samples, and yellow triangles denote observations.}\label{Fig3}
\end{figure*}

\begin{figure*}[!th]
\centerline{\includegraphics[width=\textwidth,angle=0]{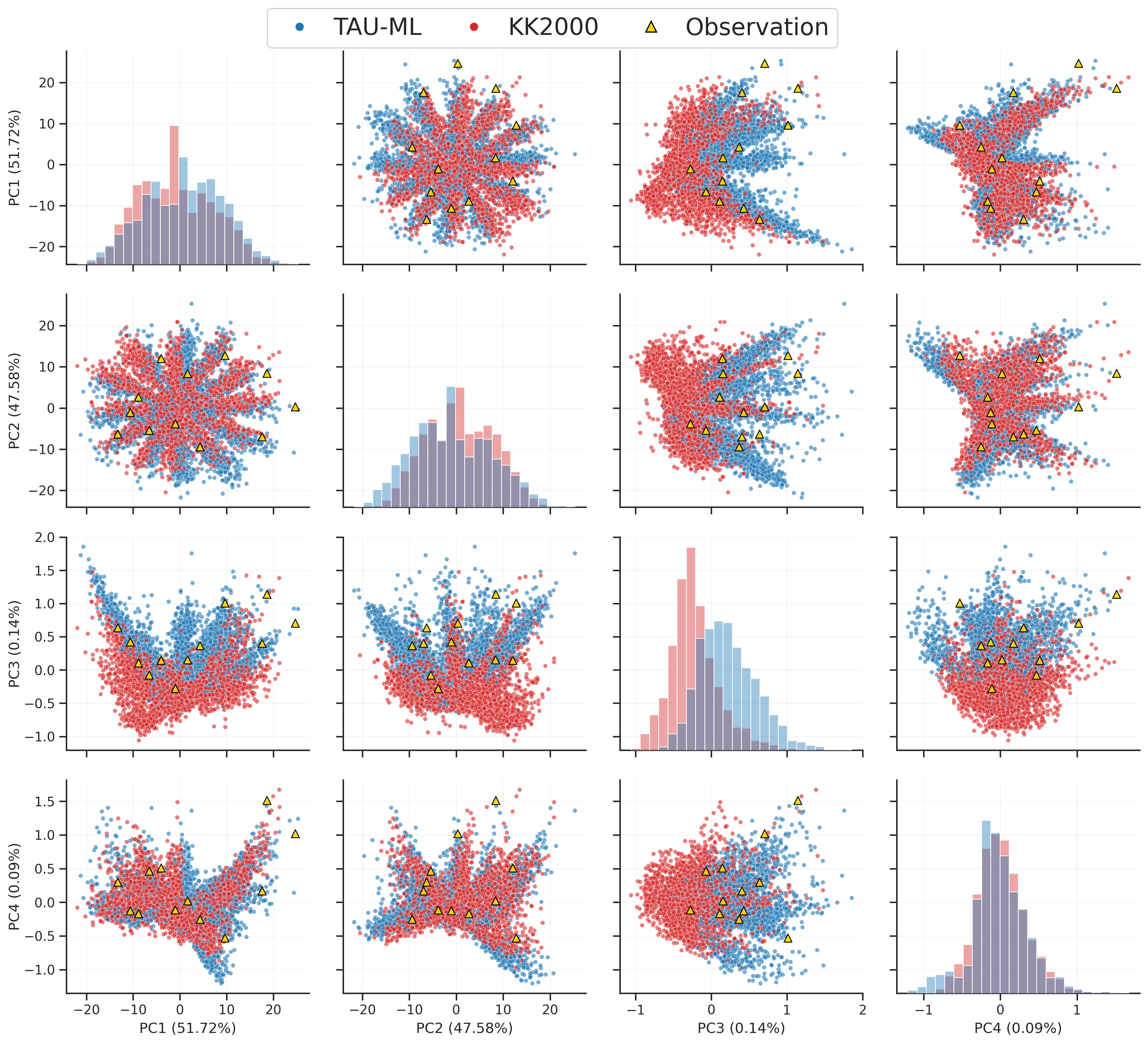}}
\caption{Same as Fig~\ref{Fig3}, but for the last layer (dense3) embeddings.}\label{Fig4}
\end{figure*}

The learned embedding space provides a compact, low-dimensional representation of the input fields. Figures~\ref{Fig3} and~\ref{Fig4} show pairwise two-dimensional (2D) PCA projections of TAU-ML, KK2000, and observation embeddings generated by the encoder and last projection layer respectively. For the encoder embeddings (Fig.~\ref{Fig3}), the PC1-PC2 projection shows a circular structure with around 12 clusters, corresponding to the seasonal cycle. Within each cluster, TAU-ML and KK2000 embeddings are generally well separated with some overlap, consistent with its high classification skill in Fig~\ref{Fig2}. The spread of TAU-ML and KK2000 embeddings within each cluster reflects ensemble variability due to parameter perturbations. Across the PCA projections, and particularly in the PC1-PC2 panel, observation embeddings are distributed along the same low-dimensional manifold as the KK2000 and TAU-ML embeddings, rather than forming an isolated cluster. This suggests that the learned representations provide a common embedding space in which model outputs and observations can be compared directly. In the PC1-PC2 and PC3-PC4 projections, some observation embeddings are displaced from the corresponding TAU-ML and KK2000 clusters, indicating structural error in the model outputs relative to observations.

%The PC1-PC3 panel shows a parabolic structure that reflects variability 
%Across all panels, and particularly in the PC1–PC2 projection, the observation embeddings are distributed along the same low-dimensional manifold as the model-derived embeddings, rather than forming a distinct or isolated cluster. This indicates that the learned representation places observations within the same underlying feature space as the models.
%which is necessary if the representation is to support a common comparison among model configurations and observations. The approximately circular structure in PC1--PC2, with clusters corresponding to the 12 calendar months, indicates that the encoder retains the seasonal organization of the input fields. Because PC1 and PC2 explain only about 48 percent of the encoder variance, the parabolic structure in the PC1--PC3 and PC2--PC3 projections likely reflects additional cloud-radiative variability that is not captured by the leading seasonal plane. 

The last layer embeddings (Fig.~\ref{Fig4}) show a more compact structure than the encoder embeddings. The first two PCs account for nearly all of the embedding variance, while PC3 and PC4 together explain only 0.23\%. The PC1-PC2 projection forms a radial pattern with around 12 arms extending from the center, corresponding to the seasonal cycle. The TAU-ML and KK2000 embeddings are more densely mixed near the center, where embeddings from adjacent months are closer to each another. This suggests that the contrastive loss preserves similarity within positive pairs of samples from adjacent months. In this projection, TAU-ML and KK2000 embeddings overlap substantially, and observation embeddings fall on the same manifold. In contrast, PC3 shows a clear separation between TAU-ML and KK2000 despite accounting for a small fraction of the total variance. This separation is evident in the PC1-PC3 and PC2-PC3 projections, as well as in the PC3 histogram, where TAU-ML projections are shifted towards higher values and KK2000 projections toward lower values. These behaviors are consistent with the effect of the contrastive loss, which emphasizes features that distinguish negative pairs of KK2000 and TAU-ML samples while preserving similarity within positive pairs of samples at adjacent months. 
%optional: add monthly variation of last layer embedding to support the last sentence arguement.

In the following analysis, we focus on the encoder embeddings because they preserve a more general representation of the input fields, whereas the last layer embeddings emphasize discriminative and similarity features that are optimized by the contrastive loss.

\begin{figure*}[!th]
\centerline{\includegraphics[width=\textwidth,angle=0]{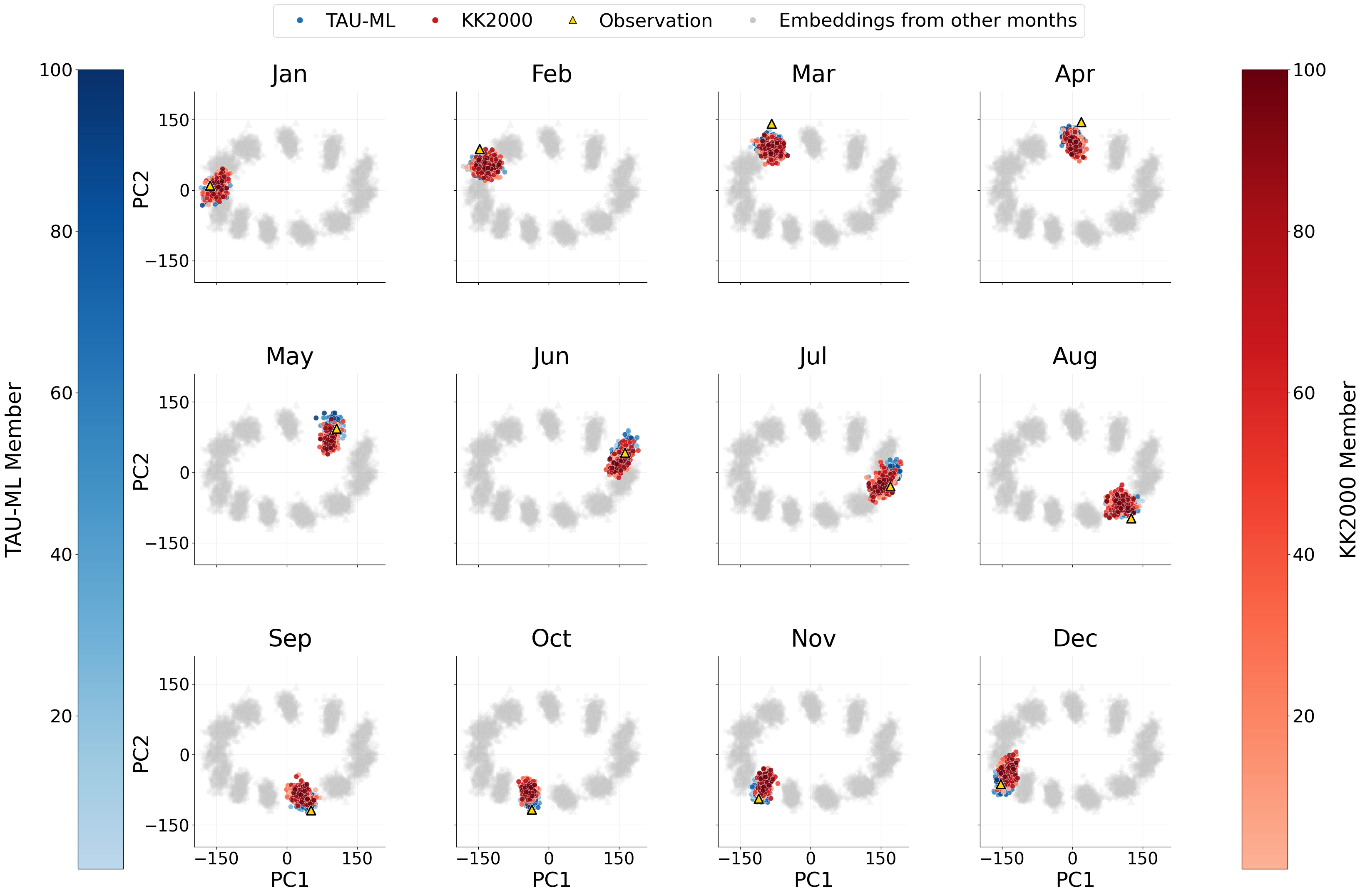}}
\caption{Monthly two-dimensional PCA projections of the learned encoder embeddings for TAU-ML, KK2000, and observations. Blue points represent TAU-ML outputs, red points represent KK2000 outputs, and yellow triangles represent observations. The colorbars indicate ensemble member index for TAU-ML and KK2000.}\label{Fig5}
\end{figure*}
The monthly 2D PCA projections of the encoder embeddings in Figure~\ref{Fig5} further show that the learned representation captures physically meaningful variability while retaining temporal coherency. Embeddings from each month form a distinct cluster in the PC1-PC2 space, and these clusters rotate clockwise with month, reflecting the seasonal cycle. Within each monthly cluster, the TAU-ML and KK2000 samples generally occupy similar regions, with their spread representing ensemble variability arising from parameter perturbations. The observation embeddings generally follow the same seasonal progression as the model embeddings, suggesting that the encoder places observations and model outputs on a shared low-dimensional manifold. However, observations are displaced from the TAU-ML and KK2000 clusters in several months in spring and autumn, including March, April, September, and October. These offsets indicate months in which CAM exhibits structural error relative to observations. These results suggest that the encoder embeddings capture ensemble spread associated with parametric uncertainty, structural differences relative to observations, and seasonal variability in the input fields.

\subsection{Embedding distance analysis}
\begin{figure*}[!th]
\centerline{\includegraphics[width=\textwidth,angle=0]{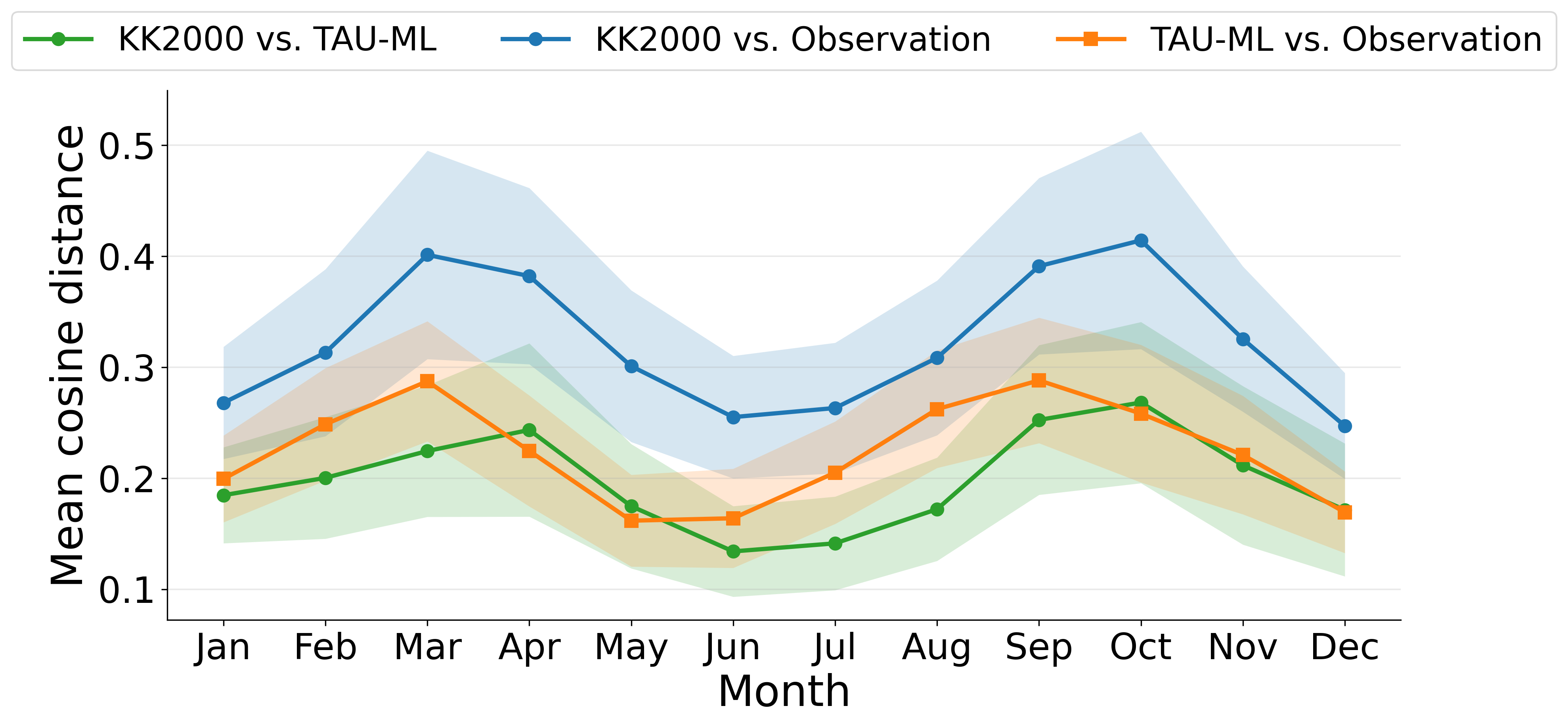}}
\caption{Monthly variation in mean cosine distance among encoder embeddings of monthly climatology for TAU-ML, KK2000, and observations. Shading represents standard deviation of the ensemble.}\label{Fig6}
\end{figure*}
To better understand differences in the embedding space, we analyzed the cosine distance among monthly climatology embedding for TAU-ML, KK2000, and observations (Fig.~\ref{Fig6}). The monthly mean distances show a clear seasonal dependence. The distances between KK2000 and TAU-ML are generally smaller than the distances between either KK2000 or TAU-ML and observations. This indicates that differences between the two warm rain microphysics in CAM are smaller than the overall discrepancy between CAM and observations.

The TAU-ML and observation distances are consistently lower than the KK2000 and observation distances, suggesting that TAU-ML has a smaller bias relative to observations than KK2000. This improvement likely results from more complex representations of cloud and rain water in TAU-ML. The KK2000 and observation distances are largest in spring and autumn, especially in March, April, September, and October, indicating seasons when KK2000 differs most from observations. TAU-ML also shows relatively large distances from observations during these months, suggesting persistent seasonal model biases. In addition, the KK2000 and observation distances exhibit larger ensemble spread than the TAU-ML and observation distances, indicating that KK2000 biases are more sensitive to physical parameters than TAU-ML.

\begin{figure*}[!th]
\centerline{\includegraphics[width=\textwidth,angle=0]{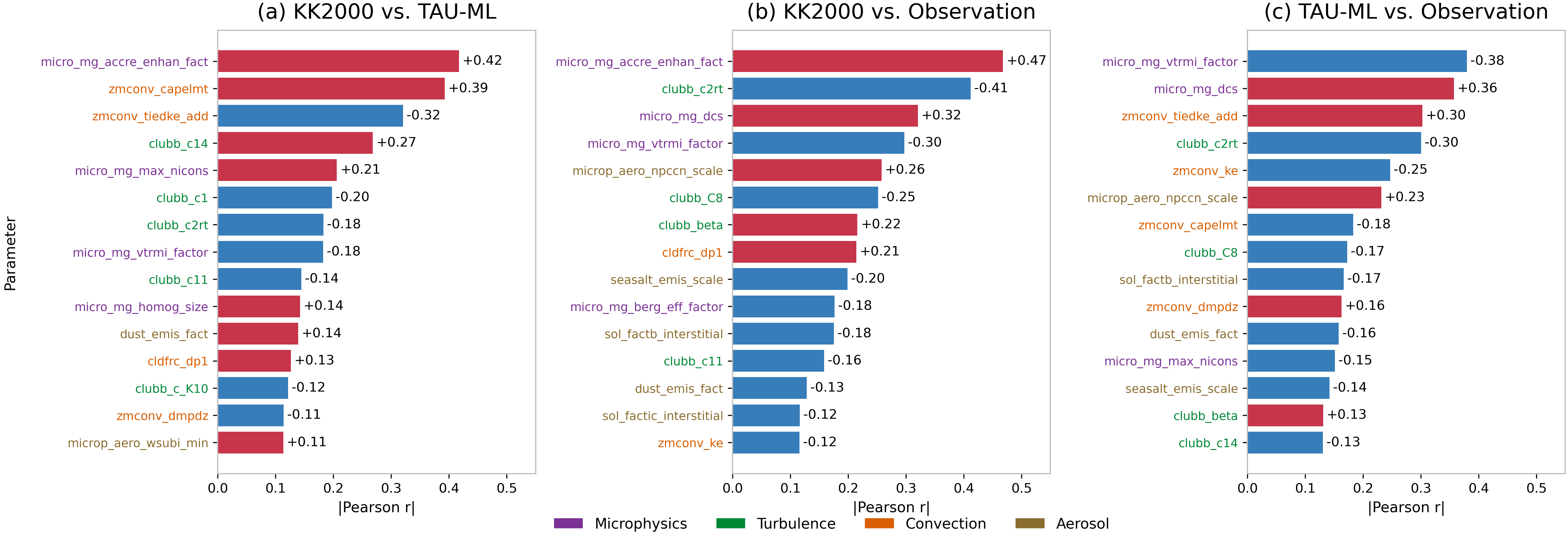}}
\caption{Top-15 parameter importance for latent distances of KK2000 a TAU-ML, KK2000 versus observations, and TAU-ML versus observations. Red bars indicate positive correlation while blue bars indicate negative correlation.}\label{Fig7}
\end{figure*}
%couples of capelmt accretion effct, don't speculative
Figure ~\ref{Fig7} shows the top-15 parameter importance for embedding distances of KK2000 versus TAU-ML, KK2000 versus observations, and TAU-ML vs observations. For KK2000 versus TAU-ML, the top-3 most important parameters are the accretion enhancement factor (\texttt{micro\_mg\_accre\_enhan\_fact}), the convective available potential energy (CAPE) triggering threshold for deep convection (\texttt{zmconv\_capelmt}), and convective parcel temperature perturbation \texttt{zmconv\_tiedke\_add}. This indicates the major difference between KK2000 and TAU-ML are their different responses to the accretion and deep convection triggering processes, consistent with the major difference between the two warm rain microphysics schemes.
%increasing this makes CLUBB behave closer to complete or no cloudiness (no variance) and brightens clouds.
%check clubb c2rt boundary layer
For KK2000 versus observations, the top-3 most important parameters are accretion (\texttt{micro\_mg\_accre\_enhan\_fact}), the damping of scalar variances for liquid water (\texttt{clubb\_c2rt}), and the autoconversion size threshold of cloud ice to snow (\texttt{micro\_mg\_dcs}). This indicates the major biases of KK2000 are in the representation of autoconversion and accretion process. For TAU-ML versus observations, the top-3 most important parameters are the ice fall speed scaling (\texttt{micro\_mg\_vtrmi\_factor}), the autoconversion size threshold of cloud ice to snow (\texttt{microp\_mg\_dcs}), and convective parcel temperature perturbation (\texttt{zmconv\_tiedke\_add}). This suggests that the TAU-ML biases results from the autoconversion and deep convection process.

\subsection{Embedding distance explanation}
%global vs regional
\begin{figure*}[!th]
\centerline{\includegraphics[width=\textwidth,angle=0]{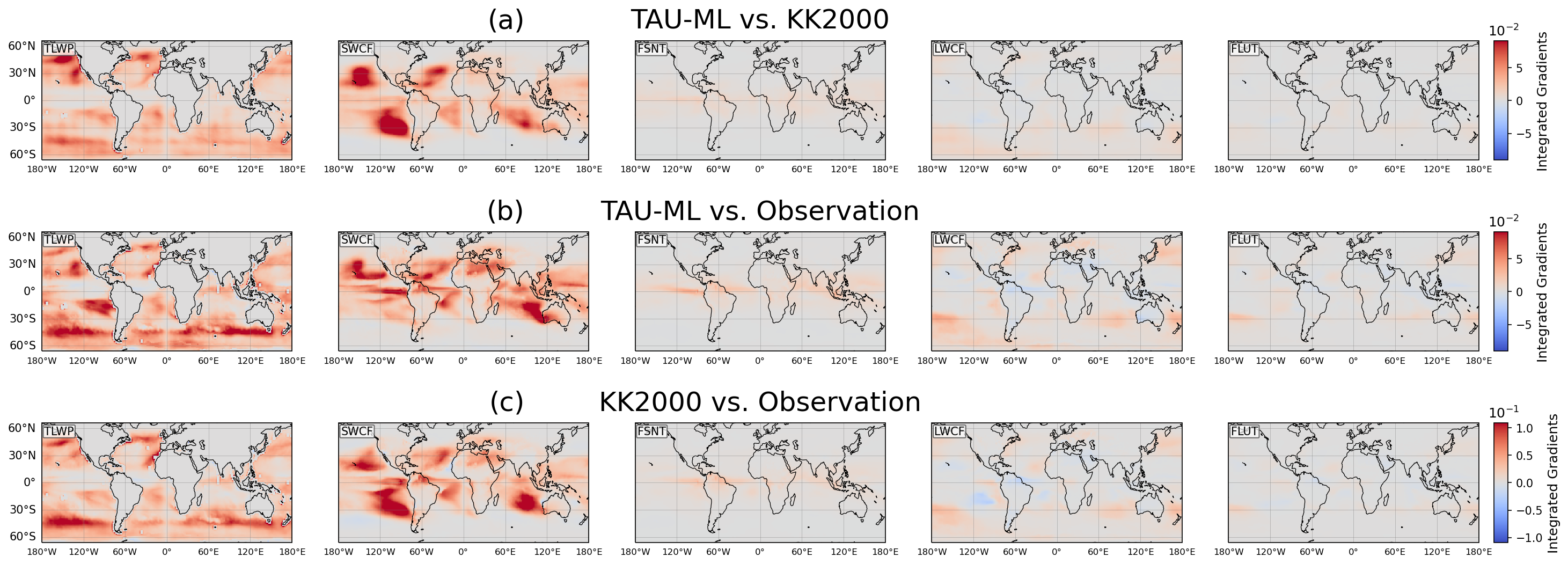}}
\caption{Mean integrated gradient attribution maps for embedding distances among TAU-ML, KK2000, and observations across months and members. Rows show (a) TAU-ML versus KK2000, (b) TAU-ML versus observations, and (c) KK2000 versus observations. Columns show TLWP, SWCF, FSNT, LWCF, and FLUT. Positive and negative values indicate regions where each input variable contributes to the latent distance between the datasets.}\label{Fig8}
\end{figure*}
To explain the embedding distances, we applied integrated gradients to identify the variables and geographic regions that contribute to the separation among TAU-ML, KK2000, and observations in the learned embedding space. Fig.~\ref{Fig8} shows the mean attribution maps for pairwise embedding distances averaged across months and ensemble members. The attributions are dominated by TLWP and SWCF, indicating that cloud processes and their radiative effects dominate the embedding space differences among the datasets.

For the TAU-ML and KK2000 comparison (Fig.~\ref{Fig8}a), the TLWP attributions are concentrated over the northeast Pacific off western North America, the North Atlantic storm track region, and the Southern Hemisphere storm track regions, with weaker signals over the subtropical South Pacific Ocean off the western South America. The SWCF attributions show strong localized contributions over the subtropical North Pacific, the southeastern Pacific off South America, the North Atlantic, and the southern Indian Ocean. FSNT only shows weak signals in the tropical ocean. LWCF also shows weak attributions in Southern and Northern Hemisphere storm track regions. FLUT contributions are negligible over the globe.

The attributions for TAU-ML and observation comparison (Fig.~\ref{Fig8}b) and KK2000 and observation comparison (Fig.~\ref{Fig8}c) show a generally similar structure but with stronger TLWP attributions in the Southern Pacific Ocean and Southern Hemisphere storm track regions and stronger SWCF attributions in the tropics. Overall, the attribution maps indicate that the embedding distances are dominated by cloud processes over oceans, especially in the Southern and Northern Hemisphere storm track regions and subtropical ocean regions.

%highlight gradient rather than diff
\begin{figure*}[!th]
\centerline{\includegraphics[width=\textwidth,angle=0]{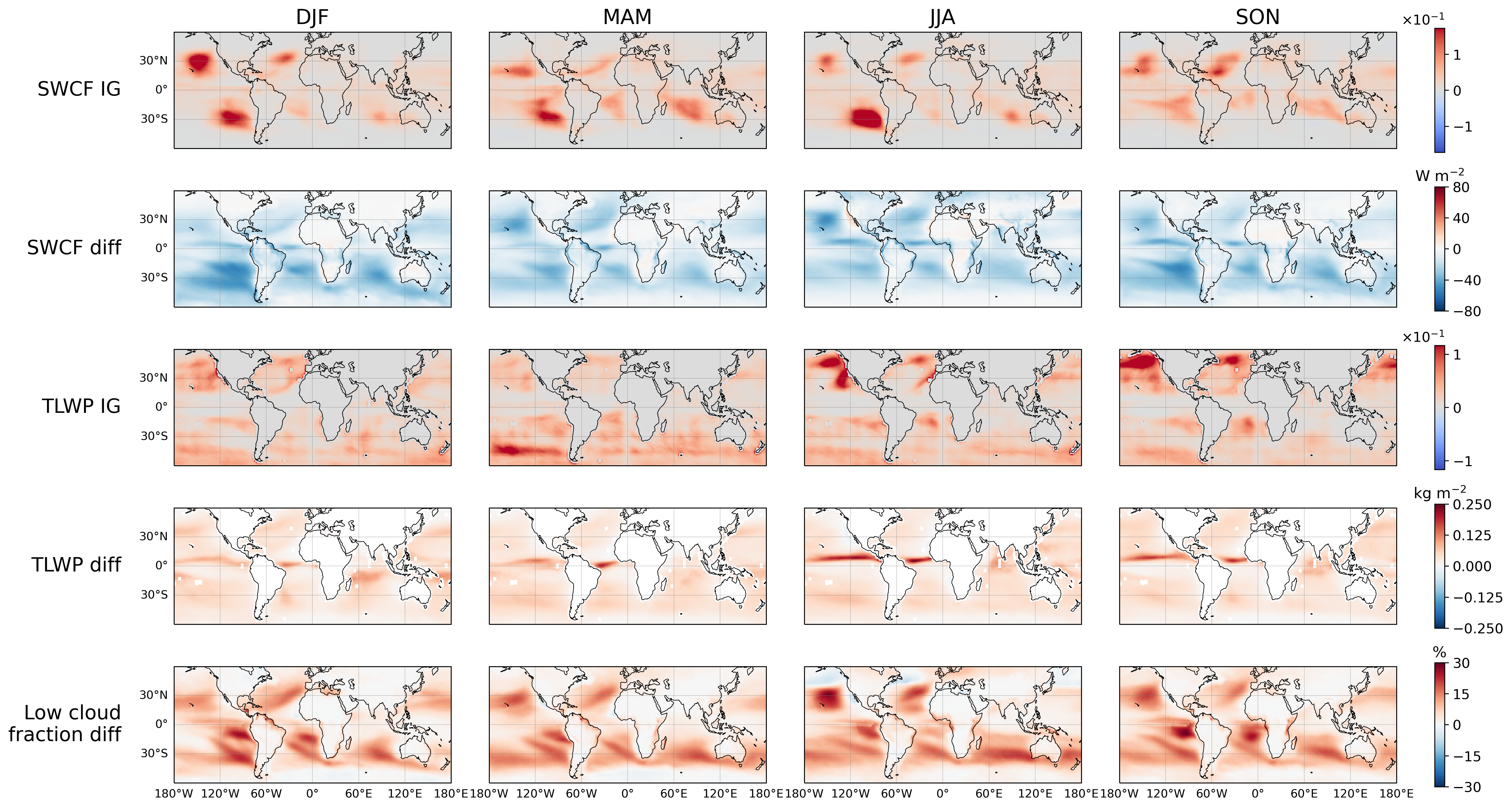}}
\caption{Seasonal integrated gradient attribution maps of SWCF and TLWP and differences of TLWP, SWCF, and low cloud fraction for the TAU-ML and KK2000 comparison. Columns show December-January-February (DJF), March-April-May (MAM), June-July-August (JJA), and September-October-November (SON). Rows show, from top to bottom, SWCF integrated gradient (IG) attribution, SWCF difference, TLWP IG attribution, TLWP difference, and low cloud fraction difference. The differences are calculated as TAU-ML minus KK2000. Positive and negative IG values indicate regions where each input variable contributes to the latent distance between the KK2000 and TAU-ML.}\label{Fig9}
\end{figure*}
%Mean differences

\begin{figure*}[!th]
\centerline{\includegraphics[width=\textwidth,angle=0]{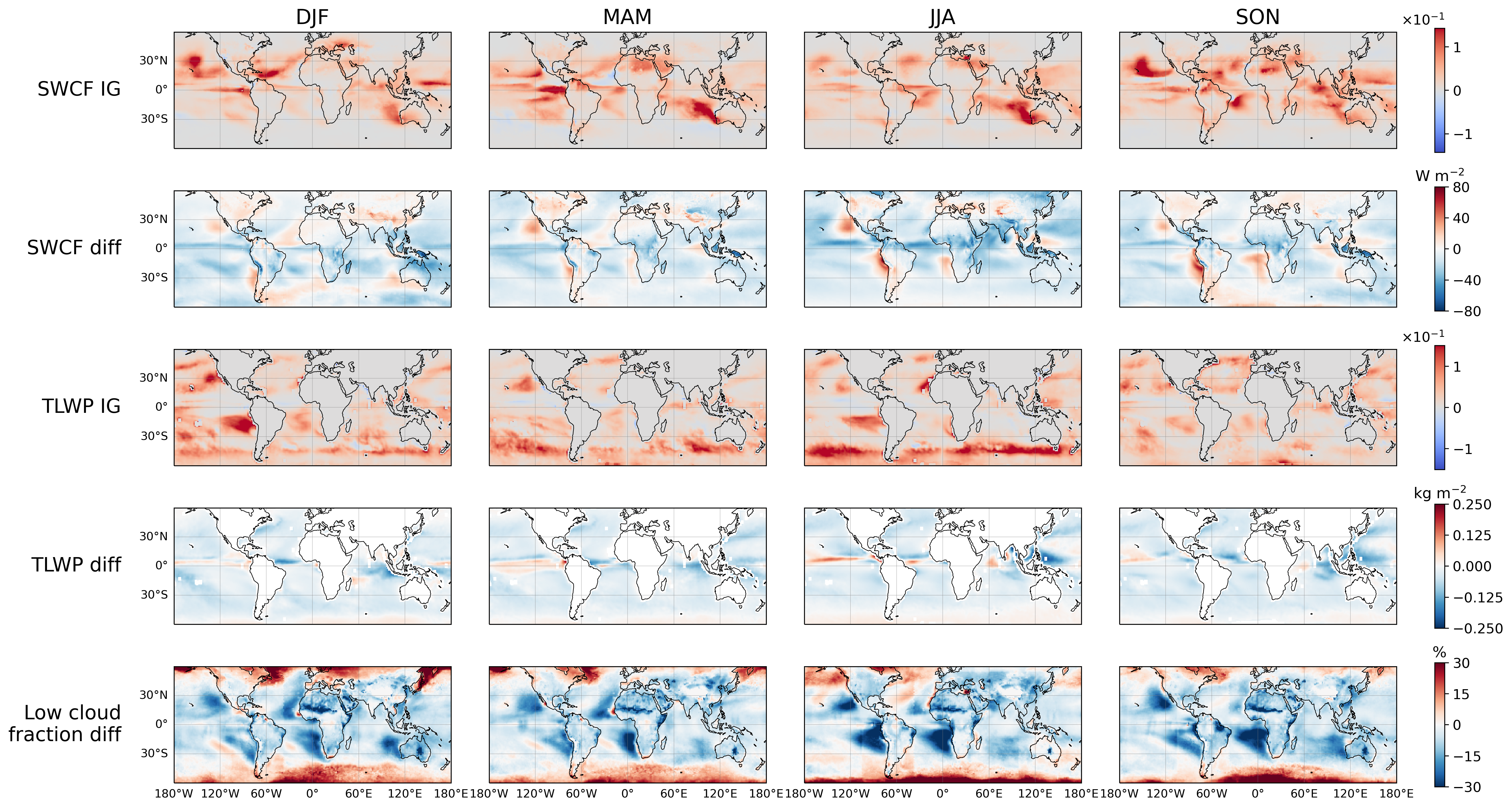}}
\caption{Same as Fig.~\ref{Fig9}, but for TAU-ML and observations comparison}\label{Fig10}
\end{figure*}

\begin{figure*}[!th]
\centerline{\includegraphics[width=\textwidth,angle=0]{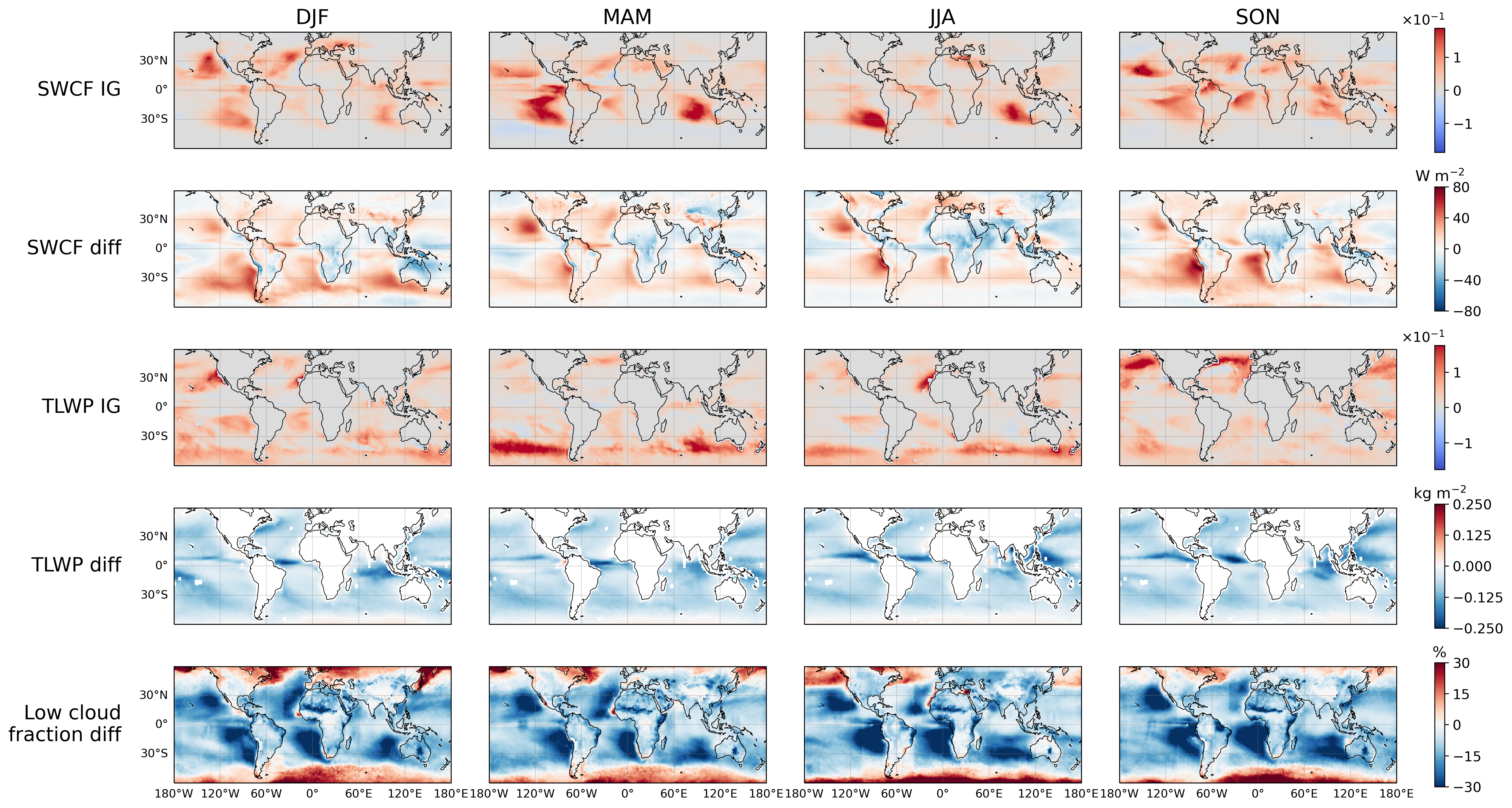}}
\caption{Same as Fig.~\ref{Fig9}, but for TAU-ML and observations comparison}\label{Fig11}
\end{figure*}

We further analyzed the seasonal variation of the attributions and their connections to cloud and radiative features. Figs.~\ref{Fig9}-\ref{Fig11} show the seasonal integrated gradient maps for SWCF and TLWP, and the seasonal differences in SWCF, TLWP, and low cloud fraction for the TAU-ML and KK2000, TAU-ML and observation, and KK2000 and observation comparisons, respectively. 

For the TAU-ML and KK2000 comparison (Fig.~\ref{Fig9}), the SWCF attributions are concentrated in subtropical low cloud regions with large low cloud fraction differences. In subtropical ocean regions, TAU-ML generates more low clouds than KK2000, leading to stronger shortwave feedback, as shown in the SWCF difference and attribution. This suggests that SWCF attributions identify regions where changes in warm rain microphysics affect SWCF through their influences on low cloud processes. The locations of the strongest signals vary seasonally with the spatial pattern of low cloud differences. The southeastern Pacific stratocumulus region shows stronger positive attribution in DJF and JJA than in the other seasons, while the northeastern Pacific signal is stronger in DJF. 

The TLWP attributions show strongest signals in the midlatitude storm track regions, especially over the North Pacific, North Atlantic, and Southern Hemisphere storm track regions. The signals in the Northern Hemisphere storm track region are stronger in JJA and SON, while the signals in the Southern storm track region are strongest in MAM. This indicates that TAU-ML and KK2000 embedding distance varies seasonally with TLWP differences in midlatitude storm track regions. The TLWP and SWCF differences over the tropical oceans only show moderate signals in TLWP and SWCF attributions. This suggests that the differences between TAU-ML and KK2000, identified by the embedding distance, are dominated by low cloud systems, directly affected by warm rain microphysics, with a smaller contribution from tropical deep convection.

For the TAU-ML and observation comparison (Fig.~\ref{Fig10}), the SWCF attributions are concentrated in subtropical low cloud regions and tropical convective regions. TAU-ML shows low cloud biases relative to observations over subtropical and tropical oceans, leading to weaker shortwave feedback from low clouds than observations. However, SWCF differences show TAU-ML generates more SWCF than observations in these regions, suggesting that there are other biases leading to negative SWCF biases while partially offset by low cloud fraction biases. The SWCF attributions identify regions with significant SWCF biases, especially over tropical oceans. This suggests that SWCF attributions identify regions affected by the biases in deep convection, a structural issue in the CAM model. The locations of the strongest signals show seasonal variations. The northeastern Pacific signal is stronger in DJF and SON, while Indian Ocean signals are stronger in MAM and JJA. Tropical ocean signals are stronger in DJF and MAM.

The TLWP attributions show strong signals in the midlatitude storm track regions and subtropical low cloud regions. The TLWP attributions show seasonal variations, consistent with the low cloud bias variation. It suggests that TLWP attributions reflect the low cloud bias. The Southern Hemisphere storm track regions show stronger signals in MAM and JJA, while the North Pacific and South Pacific signals are stronger in DJF. This suggests that the differences between TAU-ML and observations, indicated by their embedding distance, are dominated by both low cloud systems and tropical deep convection.

For the KK2000 and observation comparison (Fig.~\ref{Fig11}), the SWCF attributions are concentrated in subtropical low cloud regions and tropical convective regions. KK2000 shows widespread low cloud biases relative to observations over tropical and subtropical oceans, leading to much weaker shortwave feedback than observations, as shown in the SWCF differences and SWCF attributions. This suggests that SWCF attributions identify regions where cloud fraction biases affect shortwave radiation. The locations of the strongest signals vary seasonally with the spatial pattern of low cloud differences. The northeastern Pacific signal is stronger in DJF and SON, while the southeastern Pacific and southern Indian Ocean signals are stronger in MAM and JJA. 

The TLWP attributions show strong signals in the midlatitude storm track regions and subtropical low cloud regions. The seasonal variations of TLWP attributions in the subtropical and midlatitude regions are consistent with the low cloud fraction differences. It suggests that TLWP attributions mainly reflect the low cloud bias. The Southern Hemisphere storm track shows stronger signals in MAM and JJA, while the North Pacific and North Atlantic signals are stronger in SON. KK2000 generally underestimates TLWP relative to observations over broad tropical and subtropical ocean regions, likely due to the low cloud bias. This suggests that KK2000 and observation differences, indicated by their embedding distance, are dominated by significant low cloud biases in the subtropical ocean and storm track regions.

%marked in the plot, Microphysics groups, rename, remove micro_mg, reshape the figure
\clearpage
\begin{figure*}[!th]
\centerline{\includegraphics[width=\textwidth,angle=0]{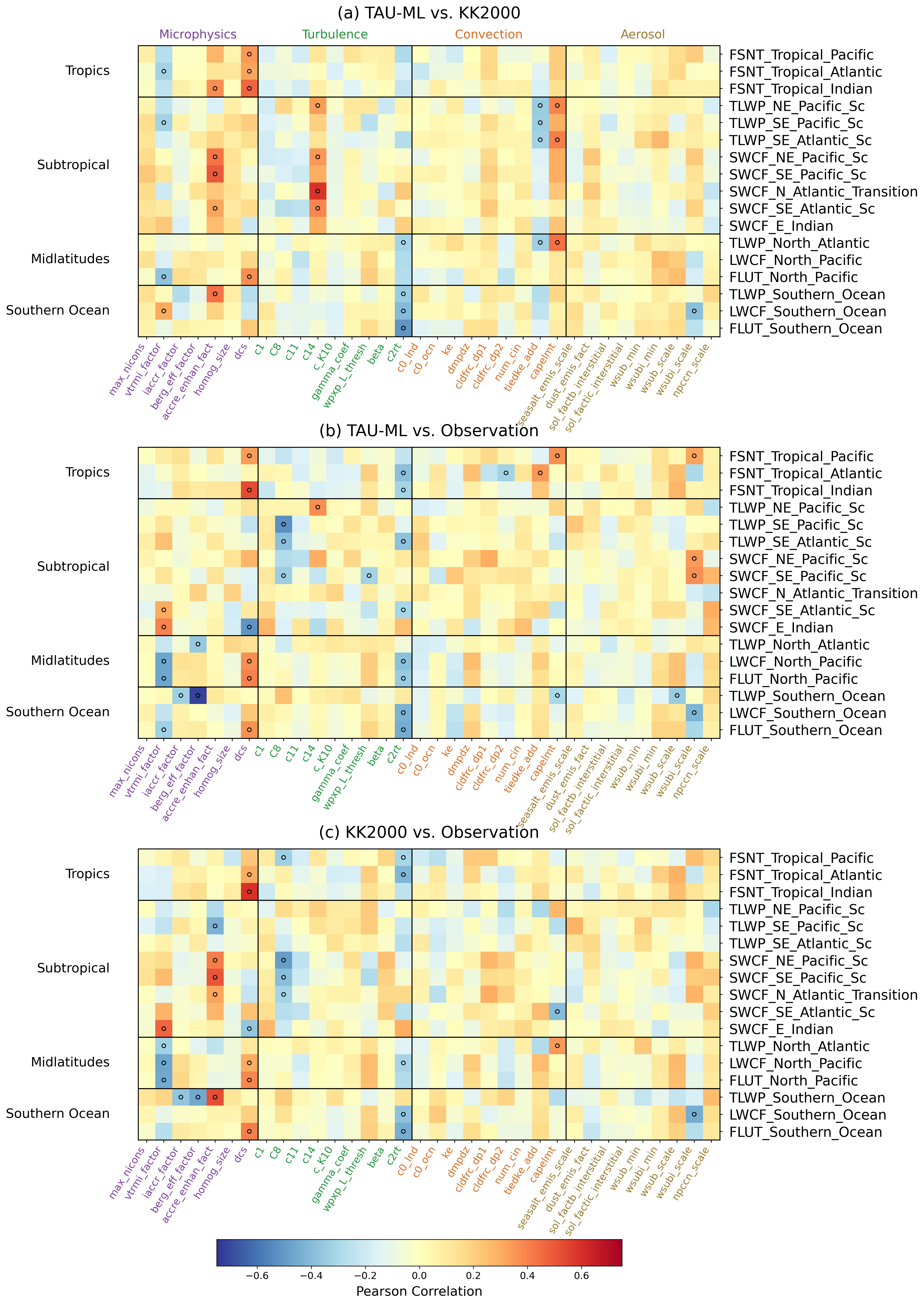}}
\caption{Pearson correlations between perturbed physics parameters and regional integrated gradient attributions for selected high-attribution variable and region pairs. Rows show selected regions for TLWP, SWCF, FSNT, LWCF, and FLUT, including marine stratocumulus, subtropical cloud-transition, ITCZ, and storm-track regions. Columns show perturbed parameters associated with cloud microphysics, CLUBB boundary layer turbulence and shallow convection, deep convection, cloud fraction, aerosol emissions, solubility, and aerosol microphysics. Panels show correlations for (a) TAU-ML versus KK2000, (b) TAU-ML versus observations, and (c) KK2000 versus observations. Colors indicate Pearson correlation coefficients between parameter values and regionally averaged integrated gradients. Open circles indicate statistically significant correlations over 0.3. Positive correlations indicate that increasing parameter values would increase the attribution in the region, while negative correlations indicate that increasing parameter values would decrease the attribution in the region.}\label{Fig12}
\end{figure*}

\subsection{Regional parameter importance}
We further examine the relationship between regional integrated gradient and the perturbed parameters. Fig.~\ref{Fig12} shows the Pearson correlations between individual parameters and the regional attributions for selected variables and regions. Positive correlations indicate that increasing parameter values would increase the attribution in the region, while negative correlations indicate that increasing parameter values would decrease the attribution in the region.

Figure ~\ref{Fig12} suggests that many parameters appear to have little impact for all selected regions and variables. Significant correlations are found in MG microphysics parameters, CLUBB turbulence parameters, and several ZM convection parameters. For the TAU-ML and KK2000 comparison (Fig.~\ref{Fig12}a), the accretion enhancement factor (\texttt{micro\allowbreak\_mg\allowbreak\_accre\allowbreak\_enhan\allowbreak\_fact}) shows strong positive correlations with TLWP in the southern Ocean storm track regions, SWCF in the subtropical stratocumulus regions, and FSNT in tropical Indian Ocean. The autoconversion size threshold of cloud ice to snow (\texttt{microp\allowbreak\_mg\_dcs}) shows positive correlation with FSNT in the tropical oceans and FLUT in the North Pacific storm track regions. The CLUBB parameter \texttt{clubb\_c14} shows positive correlations with TLWP and SWCF in the subtropical stratocumulus regions. The damping of scalar variances for liquid water (\texttt{clubb\_c2rt}) is important in the southern storm track regions. The convective available potential energy (CAPE) triggering threshold for deep convection (\texttt{zmconv\_capelmt}), and convective parcel temperature perturbation \texttt{zmconv\allowbreak\_tiedke\allowbreak\_add} have opposite impacts on TLWP in the subtropical stratocumulus regions and North Atlantic storm track region. This suggests that the TAU-ML and KK2000 differences are dominated by parameters related to low cloud processes and deep convection.

\sloppy
The TAU-ML and observation comparison (Fig.~\ref{Fig12}b) highlights the importance of microphysics and turbulence parameters. The scaling ice and snow accretion (\texttt{micro\_mg\_berg\_eff\_factor}) and Bergeron efﬁciency factor (\texttt{micro\_mg\_iaccr\_factor}) show negative correlations with TLWP in Southern and Northern storm track regions. The autoconversion size threshold of cloud ice to snow (\texttt{microp\_mg\_dcs}) shows positive correlations with FSNT in tropical oceans and LWCF and SWCF in the Northern and Southern storm track regions, and negative correlations with SWCF in the Indian Ocean. The \texttt{clubb\_C8} shows negative correlations with SWCF and TLWP in the subtropical stratocumulus regions. The damping of scalar variances for liquid water (\texttt{clubb\_c2rt}) shows negative correlations with TLWP and SWCF in the southeast Atlantic stratocumulus region, FSNT in the tropical oceans, and LWCF and FLUT in Northern and Southern storm track regions. The convective available potential energy (CAPE) triggering threshold for deep convection (\texttt{zmconv\_capelmt}), and convective parcel temperature perturbation \texttt{zmconv\_tiedke\_add} show positive correlations with FSNT in the tropical Pacific ocean. This suggests that the bias of TAU-ML is dominated by parameter associated with autoconversion and accretion processes in storm track regions and subtropical stratocumulus regions and deep convective processes in the tropical oceans.

\sloppy
The KK2000 and observations comparison (Fig.~\ref{Fig12}c) shows a similar pattern of parameter sensitivity to the TAU-ML and observation comparison. The main difference is the accretion enhancement factor(\texttt{micro\_mg\_accre\_enhan\_fact}), which is only used by KK2000 microphysics. The accretion process shows positive correlations with TLWP in the Southern ocean storm track region and SWCF in the stratocumulus region, and negative correlations with TLWP in the subtropical stratocumulus regions and Northern Hemisphere storm track regions. This suggests that the bias of KK2000 is also dominated by parameter associated with autoconversion and accretion processes in storm track regions and subtropical stratocumulus regions.

%\section{Discussion}

\section{Conclusions}
This study developed a contrastive learning framework for diagnosing parametric uncertainties and structural deficiencies among climate model simulations and observations. We applied the framework to two CAM6 perturbed parameter ensembles in which corresponding members share the same perturbed parameter settings but differ in their warm rain microphysics schemes: the default KK2000 parameterization vs. TAU-ML, a neural-network emulator of the Tel Aviv University bin microphysics model. By training on monthly TLWP, SWCF, LWCF, FSNT, FLUT, and static geographic fields, the model learned a common embedding space in which KK2000, TAU-ML, and observations can be compared using embedding distances. %We then used integrated gradients and parameter-attribution correlations to connect those latent distances back to physical variables, regions, and process parameters.

The learned embeddings captured differences and variability within the input data. The embeddings separated TAU-ML and KK2000 samples with high linear classification accuracy, indicating that the network learned robust structural differences between the two warm rain microphysics. At the same time, the encoder embeddings retained physically meaningful organization: monthly samples formed coherent clusters that followed the seasonal cycle, while the spread within each cluster reflected ensemble variability associated with perturbed parameters. Observations generally lie on the same low-dimensional manifold as the model simulations, showing that the learned representation provides a common basis for model-observation comparison. However, observations were displaced from the model clusters during the spring and autumn, indicating more systematic atmospheric model discrepancies in seasons with spatially-shifting storm tracks.

Embedding distance analysis showed that the difference between TAU-ML and KK2000 is smaller than the difference between either model configuration and observations. TAU-ML is generally closer to observations than KK2000, suggesting that the more detailed representation of warm-rain collection processes reduces the model bias. However, both configurations exhibit substantial discrepancies from observations, particularly in spring and autumn. The larger ensemble spread in the distance between KK2000 and observations further indicates that KK2000 biases are more sensitive to perturbed physical parameters than TAU-ML biases.

The latent distances are physically explainable. Integrated gradient attribution showed that TLWP and SWCF dominate the separation among TAU-ML, KK2000, and observations, while FSNT, LWCF, and especially FLUT contribute more weakly. The strongest attribution signals occur over low cloud and deep convection regions, including subtropical low cloud regions, the northeast and southeast Pacific, the North Atlantic storm track, the southern Indian Ocean, tropical Ocean, and the broader Southern Hemisphere storm track region. Seasonal attribution maps further show that these signals vary coherently with differences in TLWP, SWCF, and low cloud fraction. This indicates that the learned embedding distances are not arbitrary statistical separations, but are tied to cloud liquid water, low-cloud structure, and their shortwave radiative effects.

The comparison between TAU-ML and KK2000 is dominated by low-cloud and storm-track regimes that are directly sensitive to warm rain microphysics. TAU-ML produces more low clouds in several subtropical ocean regions, leading to shortwave cloud-forcing differences that are reflected in the SWCF attributions. TLWP attributions are strongest in midlatitude storm track regions, especially the North Pacific, North Atlantic, and Southern Hemisphere storm tracks. In contrast, model and observation comparisons show broader and stronger attributions, including both subtropical low cloud regions and tropical convective regions. This suggests that the remaining model biases involve not only warm-rain microphysics but also coupled errors in low-cloud, storm-track, and deep-convective processes.

The sensitivity analysis of parameters and attributions provides an additional link between the learned representation and model physics. Significant correlations are concentrated in cloud microphysics, CLUBB boundary layer turbulence and shallow convection, and ZM deep-convection parameters. For the TAU-ML and KK2000 comparison, accretion and deep-convection triggering parameters strongly influence the latent separation, consistent with the structural difference between the warm-rain schemes. For the model and observation comparisons, parameters related to accretion, autoconversion, ice and snow processes, boundary layer turbulence, and convective triggering control regional attributions in storm-track, stratocumulus, and tropical ocean regions. Many aerosol-emission, solubility, and dust parameters show weaker and less systematic relationships, indicating that the dominant embedding distances in this analysis are primarily controlled by cloud and precipitation process sensitivities.

Overall, these results show that contrastive learning can provide a compact and physically explainable diagnostic space for climate model evaluation. The framework captures seasonal variability, ensemble spread, structural differences between model configurations, and discrepancies between models and observations. When combined with integrated gradients, the embedding distance can be traced back to the variables and regions responsible for the separation. This makes the approach useful not only for detecting that two simulations differ, but also for identifying where and why those differences arise.

These findings have direct implications for future calibration. A scalar global error metric may identify parameter settings that appear closer to observations, but it does not necessarily reveal whether the improvement reflects physically meaningful changes or compensating errors. The embedding distance provides a multivariate metric that preserves spatial structure, seasonal dependence, and coupled cloud-radiation behavior. The attribution and parameter-correlation analyses then identify the cloud regimes and process parameters responsible for the discrepancy between model and observation. Future work should extend this framework to uncertainty-aware observational comparisons, and calibration strategies that use the learned embedding distance as an objective function.

\begin{acknowledgments}
This work was supported by NSF funding through the Learning the Earth with Artificial intelligence and Physics (LEAP) Science and Technology Center (STC; Award No. 2019625). This work was also supported by the National Science Foundation (NSF) National Center for Atmospheric Research, which is a major facility sponsored by NSF under Cooperative Agreement No. 1852977. The authors thank Dr. Michael Pritchard for contribution of ideas and William A. Manriquez for discussions on the analysis.

%#acknowledgement, CERES data were obtained from the NASA Langley Research Center CERES ordering tool at https://ceres.larc.nasa.gov/data/. LEAP, NCAR, funding sources from coauthors.
%data sources (processed). public doi, zenodo, trained, calibration code excluded.
\end{acknowledgments}

\section{Data Availability Statement}
The datasets used in this work are made available on Zenodo (\url{https://doi.org/10.5281/zenodo.22943628}). Upon acceptance of the manuscript, the repository will be officially published and a permanent DOI will be provided. 
The CERES EBAF data were obtained from the NASA Langley Research Center CERES ordering tool at \url{https://ceres.larc.nasa.gov/data/}. MAC‐LWP data are available at \url{https://doi.org/10.5067/MEASURES/MACLWPM} \cite{Elsaesser2016MACLWPM}. 
\bibliography{main}

\end{document}